\documentclass[manyauthors,nocleardouble,COMPASS]{cernepprep}
\usepackage{bm}
\usepackage{cite}  
\usepackage{lineno}  
\usepackage{booktabs}
\usepackage{multirow}
\usepackage{color}
\usepackage{physics}
\usepackage[figuresright]{rotating}

\graphicspath{ {figures/}  {figures-new/}  }

\usepackage{etex}
\usepackage{amsmath,amssymb,fleqn,url,color,xcolor,graphics,graphicx}
\usepackage{enumerate}
\usepackage{bigstrut}
\usepackage{cite}

\usepackage{indentfirst}
\usepackage{lipsum}
\usepackage{soul}
\usepackage[normalem]{ulem}  %to cross out the text

\usepackage{hyperref}
\hypersetup{
    colorlinks,
    citecolor=blue,
    filecolor=blue,
    linkcolor=blue,
    urlcolor=blue
}

\usepackage{lineno}
\newcommand{\Jpsi}{J\!/\!\psi}
\newcommand{\GeV}{\mathrm{GeV}}
\newcommand{\pt}{p_{\mathrm{T~2}J/\psi}}
\newcommand{\pbeam}{p_{\mathrm{beam}}}
\newcommand{\pz}{p_{\mathrm{L~2}\Jpsi}}
\newcommand{\xpar}{x_{\mathrm{||~2}\Jpsi}}
\newcommand{\xF}{x_{\mathrm{F}}}
\newcommand{\mpair}{M_{\mathrm{2}\Jpsi}}
\newcommand{\sigmaPt}{\sigma^{\mathrm{Pt}}}
\newcommand{\sigmap}{\sigma^{\mathrm{p}}}
\newcommand{\fbckg}{f_{\mathrm{bkg}}(x_{||~\mathrm{2}\Jpsi})}
\newcommand{\fic}{f_{\mathrm{IC}}(x_{||~\mathrm{2}\Jpsi})}
\newcommand{\fsps}{f_{\mathrm{SPS}}(x_{||~\mathrm{2}\Jpsi})}

\newcommand{\CorAuth}[1]{$^{,\#}$} 

\begin{document}

%\linenumbers
\begin{titlepage}
\EPnumber{2022--073}
%\EPdate{\today}
\EPdate{30 March 2022}

\title{Double $J/\psi$ production in pion-nucleon scattering at COMPASS}
%\thanks{A footnote to the article title}%

%\author{...}
\Collaboration{The COMPASS Collaboration}
\ShortAuthor{The COMPASS Collaboration}
\ShortTitle{Double $J/\psi$ production in pion nucleon scattering at COMPASS}

\begin{abstract}
	We present the study of the production of double $\Jpsi$ mesons using COMPASS data collected with a 190~$\GeV/c$ $\pi^-$ beam scattering off NH$_{3}$, Al and W targets.
	Kinematic distributions of the collected double $\Jpsi$ events are analysed, and the double $\Jpsi$ production cross section is estimated for each of the COMPASS targets.
	The results are compared to predictions from single- and double-parton scattering models as well as the pion intrinsic charm and the tetraquark exotic resonance hypotheses. 
	It is demonstrated that the single parton scattering production mechanism gives the dominant contribution that is sufficient to describe the data. An upper limit on the double intrinsic charm content of pion is evaluated.
	No significant signatures that could be associated with exotic tetraquarks are found in the double $\Jpsi$ mass spectrum.
\end{abstract}
\vspace*{60pt}
\begin{flushleft}

Keywords: COMPASS, charmonia production, double $\Jpsi$, single parton scattering, intrinsic charm
\end{flushleft}

\vfill
\Submitted{(to be submitted to Phys. Lett. B)}
\end{titlepage}

{
\pagestyle{empty}
%%%%%%%%%%%%%%%%%%%%%%%%%%%%%%%%%%%%%%%%%%%%%%%%%%%%%%%%%%%%%%%
%
% 2016_auththorlist.tex (default list, updated 26.10.2016)
%
%%%%%%%%%%%%%%%%%%%%%%%%%%%%%%%%%%%%%%%%%%%%%%%%%%%%%%%%%%%%%%%
\section*{The COMPASS Collaboration}
\label{app:collab}
\renewcommand\labelenumi{\textsuperscript{\theenumi}~}
\renewcommand\theenumi{\arabic{enumi}}
\begin{flushleft}
G.D.~Alexeev\Irefn{dubna}, %1
M.G.~Alexeev\Irefnn{turin_u}{turin_i},
A.~Amoroso\Irefnn{turin_u}{turin_i},
V.~Andrieux\Irefn{illinois},
V.~Anosov\Irefn{dubna}, %3
K.~Augsten\Irefn{praguectu}, %2 phd
W.~Augustyniak\Irefn{warsaw},
C.D.R.~Azevedo\Irefn{aveiro},
B.~Bade{\l}ek\Irefn{warsawu},
M.~Ball\Irefn{bonniskp},
J.~Barth\Irefn{bonniskp},
R.~Beck\Irefn{bonniskp},
Y.~Bedfer\Irefn{saclay},
J.~Bernhard\Irefn{cern},
M.~Bodlak\Irefn{praguecu},
F.~Bradamante\Irefn{triest_i},
A.~Bressan\Irefnn{triest_u}{triest_i},
V.~E.~Burtsev\Irefn{tomsk},
W.-C.~Chang\Irefn{taipei},
C.~Chatterjee\Irefn{triest_i}\Aref{A},
M.~Chiosso\Irefnn{turin_u}{turin_i},
A.~G.~Chumakov\Irefn{tomsk},
S.-U.~Chung\Irefn{munichtu}\Aref{B}\Aref{B1},
A.~Cicuttin\Irefnn{triest_i}{triest_a},
P.~M.~M.~Correia\Irefn{aveiro},
M.L.~Crespo\Irefnn{triest_i}{triest_a}
% ...
\end{flushleft}
%%%%%%%%%%%%%%%%%%%%%%%%%%%%%%%%%%%%%%%%%%%%%%%%%%%%%%%%%%%%%%%%%%%%%%%%%%%%%%%%%%%%%%%%%%%%%%%%%%%%%%%%%%%%%%%%%%%%%%%
%
% institutes
%
%%%%%%%%%%%%%%%%%%%%%%%%%%%%%%%%%%%%%%%%%%%%%%%%%%%%%%%%%%%%%%%%%%%%%%%%%%%%%%%%%%%%%%%%%%%%%%%%%%%%%%%%%%%%%%%%%%%%%%%
%\item \Idef{bielefeld}{Universit\"at Bielefeld, Fakult\"at f\"ur Physik, 33501 Bielefeld, Germany\Arefs{l}}
%\item \Idef{munichlmu}{Ludwig-Maximilians-Universit\"at M\"unchen, Department f\"ur Physik, 80799 Munich, Germany\Arefs{l}\Arefs{r}}
\begin{Authlist}
\item \Idef{aveiro}{University of Aveiro, I3N, Dept.\ of Physics, 3810-193 Aveiro, Portugal\Arefs{U}}
%\item \Idef{bochum}{Universit\"at Bochum, Institut f\"ur Experimentalphysik, 44780 Bochum, Germany\Arefs{Q}$^,$\Arefs{R}}
\item \Idef{bochum}{Universit\"at Bochum, Institut f\"ur Experimentalphysik, 44780 Bochum, Germany\Arefs{Q}}
\item \Idef{bonniskp}{Universit\"at Bonn, Helmholtz-Institut f\"ur  Strahlen- und Kernphysik, 53115 Bonn, Germany\Arefs{Q}}
\item \Idef{bonnpi}{Universit\"at Bonn, Physikalisches Institut, 53115 Bonn, Germany\Arefs{Q}}
\item \Idef{brno}{Institute of Scientific Instruments of the CAS, 61264 Brno, Czech Republic\Arefs{S}}
\item \Idef{calcutta}{Matrivani Institute of Experimental Research \& Education, Calcutta-700 030, India\Arefs{T}}
\item \Idef{dubna}{Joint Institute for Nuclear Research, 141980 Dubna, Moscow region, Russia\Arefs{T1}}
%\item \Idef{freiburg}{Universit\"at Freiburg, Physikalisches Institut, 79104 Freiburg, Germany\Arefs{Q}$^,$\Arefs{R}}
\item \Idef{freiburg}{Universit\"at Freiburg, Physikalisches Institut, 79104 Freiburg, Germany\Arefs{Q}}
\item \Idef{cern}{CERN, 1211 Geneva 23, Switzerland}
\item \Idef{liberec}{Technical University in Liberec, 46117 Liberec, Czech Republic\Arefs{S}}
\item \Idef{lisbon}{LIP, 1649-003 Lisbon, Portugal\Arefs{U}}
\item \Idef{mainz}{Universit\"at Mainz, Institut f\"ur Kernphysik, 55099 Mainz, Germany\Arefs{Q}}
\item \Idef{miyazaki}{University of Miyazaki, Miyazaki 889-2192, Japan\Arefs{V}}
\item \Idef{moscowlpi}{Lebedev Physical Institute, 119991 Moscow, Russia}
\item \Idef{munichtu}{Technische Universit\"at M\"unchen, Physik Dept., 85748 Garching, Germany\Arefs{Q}}
\item \Idef{nagoya}{Nagoya University, 464 Nagoya, Japan\Arefs{V}}
\item \Idef{praguecu}{Charles University, Faculty of Mathematics and Physics, 12116 Prague, Czech Republic\Arefs{S}}
\item \Idef{praguectu}{Czech Technical University in Prague, 16636 Prague, Czech Republic\Arefs{S}}
\item \Idef{protvino}{State Scientific Center Institute for High Energy Physics of National Research Center `Kurchatov Institute', 142281 Protvino, Russia}
%\item \Idef{saclay}{IRFU, CEA, Universit\'e Paris-Saclay, 91191 Gif-sur-Yvette, France\Arefs{R}}
\item \Idef{saclay}{IRFU, CEA, Universit\'e Paris-Saclay, 91191 Gif-sur-Yvette, France}
\item \Idef{taipei}{Academia Sinica, Institute of Physics, Taipei 11529, Taiwan\Arefs{W}}
\item \Idef{tartu}{University of Tartu, Institute of Physics, Tartu, Estonia}
\item \Idef{telaviv}{Tel Aviv University, School of Physics and Astronomy, 69978 Tel Aviv, Israel\Arefs{X}}
\item \Idef{tomsk}{Tomsk Polytechnic University, 634050 Tomsk, Russia\Arefs{Y}}
\item \Idef{turin_u}{University of Torino, Dept.\ of Physics, 10125 Torino, Italy}
\item \Idef{turin_i}{Torino Section of INFN, 10125 Torino, Italy}
\item \Idef{triest_a}{Abdus Salam ICTP, 34151 Trieste, Italy}
\item \Idef{triest_u}{University of Trieste, Dept.\ of Physics, 34127 Trieste, Italy}
\item \Idef{triest_i}{Trieste Section of INFN, 34127 Trieste, Italy}
\item \Idef{illinois}{University of Illinois at Urbana-Champaign, Dept.\ of Physics, Urbana, IL 61801-3080, USA\Arefs{Z}}
\item \Idef{warsaw}{National Centre for Nuclear Research, 02-093 Warsaw, Poland\Arefs{a} }
\item \Idef{warsawu}{University of Warsaw, Faculty of Physics, 02-093 Warsaw, Poland\Arefs{a} }
\item \Idef{warsawtu}{Warsaw University of Technology, Institute of Radioelectronics, 00-665 Warsaw, Poland\Arefs{a} }
\item \Idef{yamagata}{Yamagata University, Yamagata 992-8510, Japan\Arefs{V} }
%\item \Idef{retired}{Retired}
\end{Authlist}
%%%%%%%%%%%%%%%%%%%%%%%%%%%%%%%%%%%%%%%%%%%%%%%%%%%%%%%%%%%%%%%%%%%%%%%%%%%%%%%%%%%%%%%%%%%%%%%%%%%%%%%%%%%%%%%%%%%%%%%
%
% Notes
%
%%%%%%%%%%%%%%%%%%%%%%%%%%%%%%%%%%%%%%%%%%%%%%%%%%%%%%%%%%%%%%%%%%%%%%%%%%%%%%%%%%%%%%%%%%%%%%%%%%%%%%%%%%%%%%%%%%%%%%%
%\vspace*{-\baselineskip}
\renewcommand\theenumi{\alph{enumi}}
\begin{Authlist}
\item [{\makebox[2mm][l]{\textsuperscript{\#}}}] Corresponding authors\\
{\it E-mail addresses}: Andrei.Gridin@cern.ch, Bakur.Parsamyan@cern.ch, Fulvio.Tessarotto@cern.ch\\
%\item [{\makebox[2mm][l]{\textsuperscript{*}}}] Deceased
%\item \Adef{A}{Also at Instituto Superior T\'ecnico, Universidade de Lisboa, Lisbon, Portugal}
\item \Adef{A}{Supported by the European Union’s Horizon 2020 research and innovation programme under grant agreement STRONG – 2020 - No 824093}
\item \Adef{B}{Also at Dept.\ of Physics, Pusan National University, Busan 609-735, Republic of Korea}
\item \Adef{B1}{Also at Physics Dept., Brookhaven National Laboratory, Upton, NY 11973, USA}
\item \Adef{B2}{Present address: NISER, Centre for Medical and Radiation Physics, Bubaneswar, India}
%\item \Adef{C}{Also at Abdus Salam ICTP, 34151 Trieste, Italy}
\item \Adef{D1}{Retired from Ludwig-Maximilians-Universit\"at, 80539 M\"unchen, Germany}
\item \Adef{D}{Supported by the DFG cluster of excellence `Origin and Structure of the Universe' (www.universe-cluster.de) (Germany)}
\item \Adef{E}{Deceased}
%\fntext[E]{Supported by CERN-RFBR Grant 12-02-91500}
\item \Adef{F}{Supported by the DFG Research Training Group Programmes 1102 and 2044 (Germany)}
\item \Adef{F1}{Supported by the European Regional Development Fund under Grant No. TK133}
\item \Adef{G}{Also at Chubu University, Kasugai, Aichi 487-8501, Japan}
\item \Adef{H}{Also at Dept.\ of Physics, National Central University, 300 Jhongda Road, Jhongli 32001, Taiwan}
\item \Adef{I}{Also at KEK, 1-1 Oho, Tsukuba, Ibaraki 305-0801, Japan}
%\item \Adef{J}{Present address: Universit\"at Bonn, Physikalisches Institut, 53115 Bonn, Germany}
\item \Adef{K}{Also at Moscow Institute of Physics and Technology, Moscow Region, 141700, Russia}
\item \Adef{L}{Also at Yerevan Physics Institute, Alikhanian Br. Street, Yerevan, Armenia, 0036}
\item \Adef{M}{Also at Dept.\ of Physics, National Kaohsiung Normal University, Kaohsiung County 824, Taiwan}
\item \Adef{M1}{Supported by ANR, France with P2IO LabEx (ANR-10-LBX-0038) in the framework ``Investissements d'Avenir'' (ANR-11-IDEX-003-01)}
\item \Adef{N}{Also at Institut f\"ur Theoretische Physik, Universit\"at T\"ubingen, 72076 T\"ubingen, Germany}
%\item \Adef{N1}{Retired}
%\item \Adef{N2}{Present address: Brookhaven National Laboratory, Brookhaven, USA}
\item \Adef{O}{Also at University of Eastern Piedmont, 15100 Alessandria, Italy}
\item \Adef{O1}{Present address: RWTH Aachen University, III.\ Physikalisches Institut, 52056 Aachen, Germany}
%\item \Adef{O2}{Present address: Universit\"at Hamburg, 20146 Hamburg, Germany}
%\item \Adef{P}{Present address: Institue of Modern Physics, Chinese Academy of Sciences, Lanzhou 730000, China}
\item \Adef{Q}{Supported by BMBF - Bundesministerium f\"ur Bildung und Forschung (Germany)}
%\item \Adef{R}{Supported by FP7, HadronPhysics3, Grant 283286 (European Union)}
\item \Adef{S}{Supported by MEYS, Grants LM20150581 and LM2018104 (Czech Republic)}
\item \Adef{T}{Supported by B.~Sen fund (India)}
\item \Adef{T1}{Supported by CERN-RFBR Grant 12-02-91500}
\item \Adef{U}{Supported by FCT, Grants CERN/FIS-PAR/0007/2017 and  CERN/FIS-PAR/0022/2019 (Portugal)}
\item \Adef{V}{Supported by MEXT and JSPS, Grants 18002006, 20540299, 18540281 and 26247032, the Daiko and Yamada Foundations (Japan)}
\item \Adef{W}{Supported by the Ministry of Science and Technology (Taiwan)}
\item \Adef{X}{Supported by the Israel Academy of Sciences and Humanities (Israel)}
%\item \Adef{Y}{Supported by Tomsk Polytechnic University Competitiveness Enhancement Program (Russia)}
\item \Adef{Y}{Supported by the Tomsk Polytechnic University within the assignment of the Ministry of Science and Higher Education (Russia)}
\item \Adef{Z}{Supported by the National Science Foundation, Grant no. PHY-1506416 (USA)}
%\item \Adef{a}{Supported by NCN, Grant 2017/26/M/ST2/00498 (Poland)}
\item \Adef{a}{Supported by NCN, Grant 2020/37/B/ST2/01547 (Poland)}
\end{Authlist}

\clearpage
%\renewcommand\labelenumi{\textsuperscript{\theenumi}~}
%\renewcommand\theenumi{\arabic{enumi}}
%\begin{flushleft}
%  \input{compass_auth_empty.tex}
%\end{flushleft}
%\begin{Authlist}
%  \input{compass_inst.tex}    
%\end{Authlist}
%\vspace*{-\baselineskip}\renewcommand\theenumi{\alph{enumi}}
%\begin{Authlist}
%  \input{compass_notes_empty.tex}
%\end{Authlist}
}

\setcounter{page}{1}

%%%%%%%%%%%%%%%%%%%%%%%%%%%%%%
\section{Introduction}
\label{section:section1}
The production mechanism of heavy quarkonia is an intriguing and challenging subject in Quantum Chromodynamics (QCD). Particularly interesting is the double quarkonia production process. 
It plays an important role in the understanding of parton interactions (single- and double-parton scattering) and parton hadronisation dynamics in high-energy collisions. 

There exist several models for double quarkonia production: single-parton scattering (SPS)~\cite{Humpert:1983yj,Ecclestone:1982yt,Kartvelishvili:1983lrw}, double-parton scattering (DPS)~\cite{Halzen:1986ue}, intrinsic charm of initial hadrons (IC)~\cite{Vogt:1995tf} or 
tetraquark decay~\cite{10.1143/PTP.54.492}. 
At moderate energies of fixed-target experiments, the production of double quarkonia is expected to be driven mainly by the SPS mechanism, while the contribution of DPS would be strongly suppressed~\cite{Lansberg:2015lva}. 
While the gluon-gluon fusion channel dominates in the SPS mechanism at collider energies, 
at the lower energies of fixed-target experiments the dominant contribution comes from quark-antiquark annihilation~\cite{Humpert:1983yj,Ecclestone:1982yt,Kartvelishvili:1983lrw}. 

First measurements of double $\Jpsi$ production cross section were performed in the 1980s by the NA3 collaboration using pion (150 $\GeV/c$ and 280 $\GeV/c$) and proton (400 $\GeV/c$) beams scattering off a platinum target~\cite{Badier:1982ae,Badier:1985ri}. More recently double $\Jpsi$ production was studied at higher centre-of-mass energies by LHCb~\cite{Aaij:2011yc,Aaij:2016bqq,Aaij:2020fnh}, D0~\cite{Abazov:2014qba}, CMS~\cite{Khachatryan:2014iia} and ATLAS~\cite{Aaboud:2016fzt}.

The results obtained by the NA3 collaboration served as a basis for the development of the aforementioned heavy quarkonia production models: SPS~\cite{Humpert:1983yj,Ecclestone:1982yt,Kartvelishvili:1983lrw}, DPS~\cite{Halzen:1986ue,Koshkarev:2019crs} and IC~\cite{Vogt:1995tf}. 
The IC model assumes the presence of non-negligible Fock components with $c\bar{c}$ pairs in a hadron wave function~\cite{Brodsky:1980pb}. In this case two $\Jpsi$ mesons are produced due to hadronisation of the $\ket{\bar{u}dc\bar{c}c\bar{c}}$ Fock state -- the double intrinsic charm component of the pion. 
As discussed in Ref.~\cite{Gridin:2019nhc}, values of kinematic variables of the double $\Jpsi$ events at NA3 were published without corresponding acceptance correction~\cite{Badier:1982ae} and therefore could not be interpreted directly.
Hence, the conclusion obtained in Ref.~\cite{Vogt:1995tf} that the NA3 data on the double $\Jpsi$ production supports the intrinsic charm hypothesis must be reconsidered. 

The existence of exotic tetraquark states made of four charm quarks was first predicted in 1975~\cite{10.1143/PTP.54.492}. Their possible decays into two $\Jpsi$ mesons were discussed in Refs.~\cite{Li:1983ru,Berezhnoy:2011xy, Debastiani:2017msn,Liu:2019zuc}.
Tetraquark states can also decay into  $\Jpsi \chi_{c_{0, 1}}$ and other intermediate states producing feed-down double $\Jpsi$ final state.
In theoretical models their contribution depends on the quantum numbers of the tetraquark state and other parameters~\cite{Chen:2020xwe}.
Alternatively, the double $\Jpsi$ final state could be produced in the decay of the bottomonium states $\eta_b$ and $\chi_{b0,2}$. According to Ref.~\cite{Braaten:2000cm}, $\eta_b$ decays into two $\Jpsi$ with a probability $7 \times 10^{-4 \pm 1}$
and the predicted branching fractions for the $\chi_{b0,2}$ are lower than $10^{-4}$~\cite{Braguta:2005gw, Tanabashi:2018oca}.

Recently the LHCb collaboration reported on the observation of the X(6900) tetraquark state in the double $\Jpsi$ mass spectrum~\cite{Aaij:2020fnh}. 
This state has a Breit-Wigner line shape and statistical significance above 5$\sigma$. It can be interpreted as a four charm tetraquark state.
The broad enhancement observed in the mass interval 6.2-6.8 GeV/$c^{2}$ can be attributed to feed-down decays of heavier quarkonia or to the mixture of $cc\bar{c}\bar{c}$ tetraquark states. 

In this Letter, the analysis of double $\Jpsi$ production events collected by the COMPASS experiment in pion scattering off different nuclear targets (ammonia, aluminium and tungsten) is presented.
The obtained kinematic distributions of double $\Jpsi$ differential cross sections are discussed and compared to model predictions (SPS, DPS and IC). The integrated production cross section is estimated for each of the nuclear targets.

%%%%%%%%%%%%%%%%%%%%%%%%%%
\section{The COMPASS experiment at CERN}
\label{section:section2}
The COMPASS experiment~\cite{Abbon:2007pq,Aghasyan:2017jop,Gautheron:2010wva} is located at the M2 beam line of the CERN Super Proton Synchrotron. 
The data used in the present analysis were collected in 2015 and 2018, using a 190~$\GeV/c$ $\pi^{-}$ beam scattering off NH$_{3}$, Al and W targets, positioned along the beam line. The NH$_3$ target consisted of two 55 cm long cylindrical cells, separated by a 20 cm gap. The cells were polarised transversely in opposite directions. Possible target polarisation effects were canceled by combining data with opposite polarisation orientations. 
A 240 cm long hadron absorber consisting of alumina blocks with a central tungsten core acting as a beam dump was installed downstream of the NH$_3$ target. 
The purpose of the hadron absorber was to strongly reduce the flux of secondary hadrons, which may decay into muons.
For the present measurement the most upstream part (10 cm) of the tungsten core was used as a heavy nuclear target.
An additional 7 cm long aluminium target was placed between the ammonia target and the tungsten core of the absorber. Outgoing charged particles were detected downstream of the absorber by a set of tracking detectors in the two-stage spectrometer. In each stage, muon identification was accomplished by muon filters, which included tracker stations separated by a hadron absorber layer. 
The trigger required the hit pattern of several hodoscope planes to be consistent with at least two muon candidate tracks originating from the targets. 
For more details see Refs.~\cite{Abbon:2007pq,Aghasyan:2017jop,Gautheron:2010wva}.

%%%%%%%%%%%%%%%%%%%%%%%%%%
\section{Selection criteria}
\label{section:section3}
In the analysis presented in this Letter, the reaction
\begin{equation}
	\label{reaction4}
	\pi^- + A \to \Jpsi~+\Jpsi+~X \to (\mu^+\mu^-)+(\mu^+\mu^-)+~X
\end{equation}
is studied. 
Events with associated incoming pion and at least two positive and two negative outgoing muon tracks originating from a primary vertex reconstructed in one of the target volumes are selected as double $\Jpsi$ candidates.
In order to minimize the effect of secondary interactions in tungsten, only events from the first 10 cm of the tungsten core are used in the analysis. 
Tracks crossing more than 30 radiation lengths of material along the spectrometer are  identified as muons.
Due to energy losses in the material of the hadron absorber and muon filters, only muons with momentum higher than 10~$\GeV/c$ could be efficiently identified.
Negative muons with momentum above 100 $\GeV/c$ produced at small angles are rejected in order to remove the strong contamination from beam pion decays. Additionally, the momentum of the four muon system is required to be smaller than 190~$\GeV/c$.
For each four-muon event, the four possible dimuon invariant mass combinations ($m_{\mu^+_1\mu^-_1}, m_{\mu^+_2\mu^-_2}, m_{\mu^+_1 \mu^-_2}, m_{\mu^+_2\mu^-_1}$) are constructed, out of which the two possible double $\Jpsi$ candidates are selected.

\begin{figure}[h!]
	\center{\includegraphics[width=0.45\linewidth]{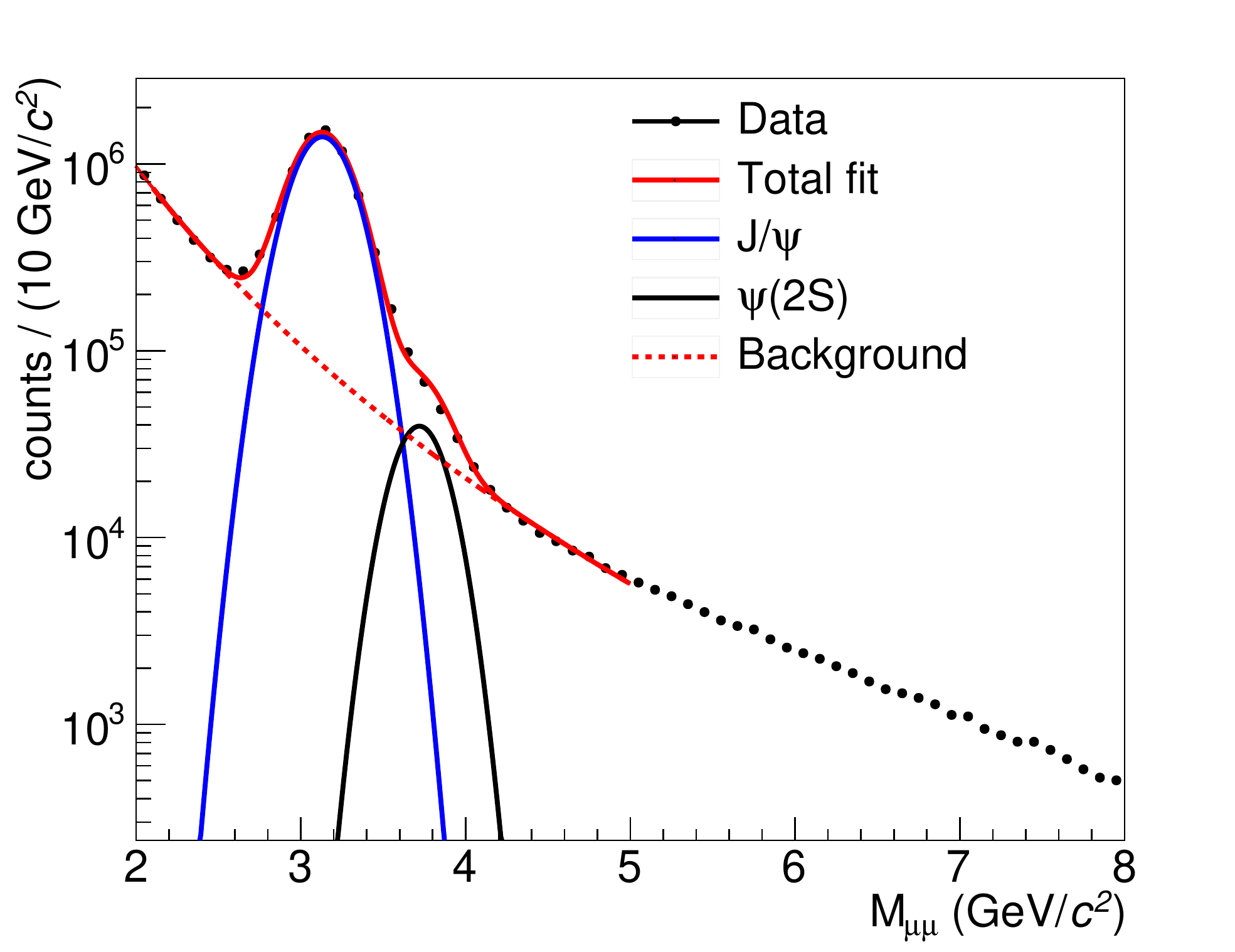}}
	\caption{\label{fig:fig1}
		Dimuon invariant mass distribution for the NH$_3$ target.}
\end{figure}

In order to estimate the double $\Jpsi$ production cross section, the results obtained for the semi-inclusive $\Jpsi$ production are needed. 
The dimuon mass distribution for events containing at least one $\mu^{+}\mu^{-}$ pair originating from the NH$_3$ target is presented in Fig.~\ref{fig:fig1}. In the mass spectrum the $\Jpsi$ peak and a shoulder from the $\psi(2S)$ resonance are clearly distinguishable.
The position and the width of the $\Jpsi$ peak are estimated by fitting a sum of two Gaussian functions (describing $\Jpsi$ and $\psi(2S)$) and a function $c_1 \cdot e^{-M_{\mu^{+}\mu^{-}} \cdot c_2} + c_3 \cdot M_{\mu^{+}\mu^{-}}^{~c_4}$ (describing the background) to the mass spectrum in the range 2.0--5.0 $\GeV/c^{2}$.
The obtained values for the peak positions, $M_{\Jpsi}$, and Gaussian width, $\Delta_{\Jpsi}$, for all the targets are presented in Tab.~\ref{tab:jpsi}. 

\begin{table}[htp]
	\caption{Single $\Jpsi$ mass and Gaussian width for all the targets.}
	\label{tab:jpsi}
	\begin{center}
		\begin{tabular}{|c|c|c|c|}
			\hline
			& NH$_3$ & Al & W \\
			\hline
			$M_{\Jpsi}$, $\GeV/c^{2}$ & 3.141 $\pm$ 0.009 & 3.138 $\pm$ 0.010 & 3.078 $\pm$ 0.009\\
			$\Delta_{\Jpsi}$, $\GeV/c^{2}$ & 0.182 $\pm$ 0.008 & 0.202 $\pm$ 0.009 & 0.299 $\pm$ 0.011 \\
			\hline
		\end{tabular}
	\end{center}
\end{table}%

\begin{figure}[h]
	\begin{center}
		\includegraphics[width=190px]{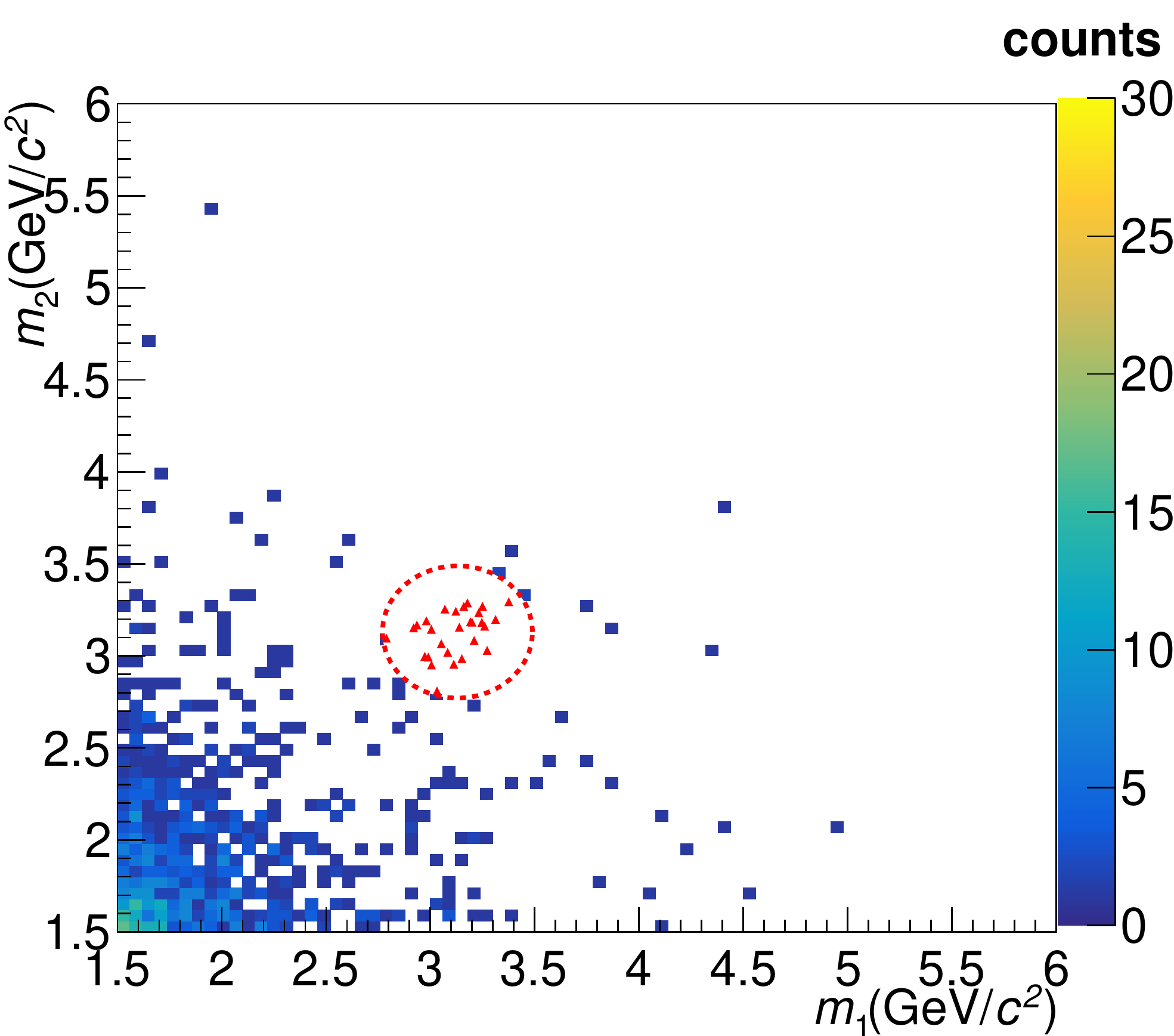}
		\includegraphics[width=220px]{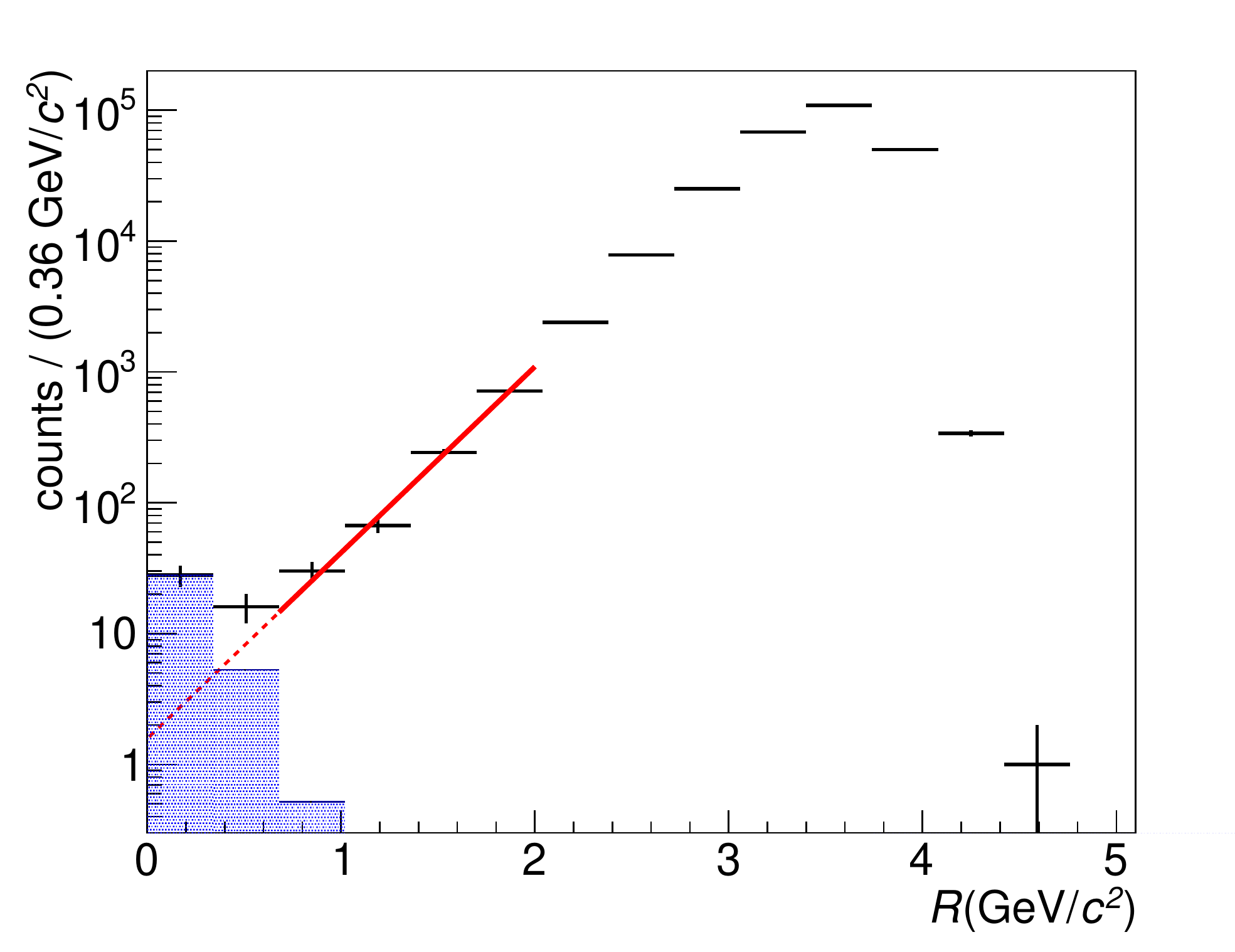}\\
		(a)\hspace{0.4\textwidth}(b)$\quad$ 
	\end{center}
	\caption{\label{fig:fig2}
		(a) The distribution of the two mass combinations, $m_1$ and $m_2$, for events with two positive and two negative muons in the final state reconstructed in the ammonia target. The selected double $\Jpsi$ candidates are shown in the red circle. (b) Distribution for the value $R$ for the real data (black) and for Monte-Carlo events of the double $\Jpsi$ production (blue). The exponential fit to the data is shown in red. 
	}
\end{figure}

Figure~\ref{fig:fig2}(a) presents the correlation of the two dimuon masses, $m_1$ and $m_2$, for double $\Jpsi$ candidates produced in the ammonia target. Here $m_1$ denotes $m_{\mu^+_1\mu^-_1}$ or $m_{\mu^+_1 \mu^-_2}$, while $m_2$ refers correspondingly to $m_{\mu^+_2\mu^-_2}$ or  $m_{\mu^+_2\mu^-_1}$.
The red circle illustrates a circular cut with radius of 2$\Delta_{\Jpsi}$ that is applied to select double $J/\psi$ candidates (marked in red). 
Events with both dimuon pair combinations passing the double $\Jpsi$ selection (one event originated in the ammonia target and five events in tungsten) are rejected. This selection does not introduce an additional systematic uncertainty, as it is used also in Monte-Carlo for the acceptance calculation.
The aforementioned selection criteria are passed by 28 events originating from NH$_3$, 2 from Al and 13 from W targets.

The nominal $\Jpsi$ mass~\cite{Tanabashi:2018oca} is assigned to the selected dimuons.
Single $\Jpsi$ candidates are required to satisfy the condition $\xF = 2p^{*}_{L} / \sqrt{s} > 0$ in order to avoid the region with low acceptance.  
Here $p^{*}_{L}$ is the longitudinal momentum of the $\Jpsi$ candidate in the centre-of-mass system. For double $\Jpsi$ candidates the same selection is applied for each $\Jpsi$ in the pair.
The complete information for each selected double $\Jpsi$ event produced in the ammonia target is available on HEPData~\cite{Maguire:2017ypu}.

In order to estimate the fraction of continuum background present under the signal region shown in Fig.~\ref{fig:fig2}(a), the distributions of radial distance $R=\sqrt{(m_1-M_{\Jpsi})^2+(m_2-M_{\Jpsi})^2}$ for each target are analysed. Here, originally measured invariant masses $m_1$ and $m_2$ are used.
The $R$ distribution for the ammonia target is shown in Fig.~\ref{fig:fig2}(b) in black. The bin width corresponds to 2$\Delta_{\Jpsi}$ such that the signal is almost entirely concentrated in the first bin of the histogram. The blue histogram represents the expected double $\Jpsi$ contribution evaluated from a Monte-Carlo simulation described in Section~\ref{section:section4}. An exponential curve is fitted to the experimental distribution 
in the range of $R$ from $4\Delta_{\Jpsi}$
to 2 $\GeV/c^2$, 
where the signal is negligible.
The background contribution in the signal region is estimated from the extrapolation of the fitted curve to the $R=0$.
After the background contributions were subtracted, the number of double $\Jpsi$ signal events $N_{2\Jpsi}$ for NH$_3$, Al and W is estimated to be: 25.1$\pm$0.5, 0.6$\pm$0.4 and 4.5$\pm$2.0, respectively.
The background for Al and W is larger than the corresponding signal.
The possible contribution of double $\Jpsi$ events produced from the decay of $B\bar{B}$ pairs is estimated to be small and is neglected.
The estimated number of signal and background events in the NH$_3$, Al and W samples are presented in Tab.~\ref{tab:Crosssections}. 

%%%%%%%%%%%%%%%%%%%%%%%%%%%%%%
\section{Results}
\label{section:section4}
\begin{figure}[h]
	\begin{center}
		\includegraphics[width=220px]{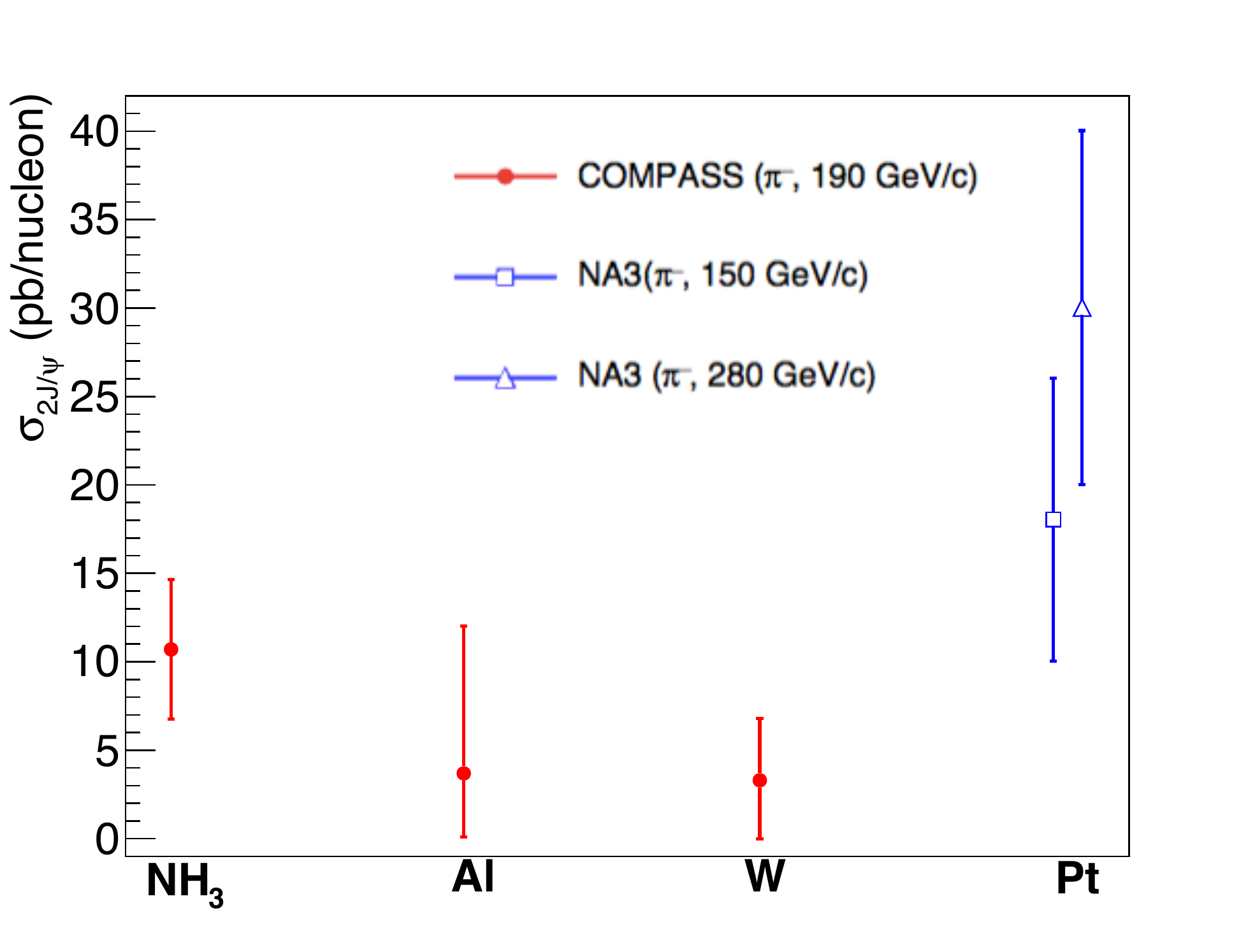}
	\end{center}
	\caption{\label{fig:fig3}
		The double $\Jpsi$ production cross section per nucleon as measured by COMPASS (black) and NA3~\cite{Badier:1982ae} (blue) in $\pi^{-}$ scattering off different targets.
	}
\end{figure}

In order to evaluate the absolute normalisation of the double $\Jpsi$ yield in the COMPASS data the NA3 single $\Jpsi$ results~\cite{Badier:1983dg} are used.
The single $\Jpsi$ production cross section was measured by NA3 using 200 GeV/c $\pi^{-}$ beam with proton target ($\sigmap_{\Jpsi} \times BR(\Jpsi \to \mu\mu)$ = 6.3 $\pm$ 0.8 nb) and with platinum target ($\sigmaPt_{\Jpsi}\times BR(\Jpsi \to \mu\mu) $ = 960 $\pm$ 150 nb). In COMPASS analysis the first value is used for an estimation of double $\Jpsi$ cross section on NH$_{3}$ and Al targets and the second value is used for the tungsten. The ratio $\sigma_{2\Jpsi}/ \sigma_{\Jpsi}$ for each target is given by the equation:
\begin{equation}
	\frac{\sigma_{2\Jpsi}}{\sigma_{\Jpsi}}=\frac{1}{BR(\Jpsi\to \mu\mu)} \times \frac{N_{2\Jpsi}}{N_{\Jpsi}}\times \frac{A_{\Jpsi}}{A_{2\Jpsi}}.
	\label{sect}
\end{equation}
Here, $\sigma_{\Jpsi}$ ($\sigma_{2\Jpsi}$) is the $\Jpsi$ (2$\Jpsi$) production cross section per nucleon for each target, while $N_{\Jpsi}$ is the number of signal $\Jpsi$ events obtained from the fit shown in Fig.~\ref{fig:fig1}. The acceptance $A_{\Jpsi}$ ($A_{2\Jpsi}$) for single (double) $\Jpsi$ events is averaged over the kinematic range of the selected samples.
The quantity $BR(\Jpsi\to \mu\mu)$ is the branching fraction of the $\Jpsi$ decay into two muons that is equal to $0.05961 \pm 0.00033$~\cite{Tanabashi:2018oca}. A Monte-Carlo simulation is performed to estimate the acceptances.
The HELAC-Onia package~\cite{Shao:2012iz,Shao:2015vga} is used to generate the hard processes (both $q\bar{q}$ annihilation and $gg$ fusion) of the double $\Jpsi$ production according to the SPS mechanism. It is assumed that the contribution from $q\bar{q}$ is two times larger than that from $gg$~\cite{Humpert:1983yj,Ecclestone:1982yt}.
The single $\Jpsi$ production is simulated with Pythia 8~\cite{Sjostrand:2014zea}.
The obtained ratio of double to single $\Jpsi$ cross sections for the ammonia target is 

\begin{equation}
	\sigma_{2\Jpsi} / \sigma_{\Jpsi} = (1.02 \pm 0.22_{stat} \pm 0.27_{syst})\cdot 10^{-4}, 
\end{equation}
which is compatible with the result reported by NA3~\cite{Badier:1982ae}.
The numerical values used in Eq.~(\ref{sect}), as well as the obtained results for the double $\Jpsi$ production cross section for each target, are listed in Tab.~\ref{tab:Crosssections}. 
In Fig.~\ref{fig:fig3} the COMPASS results are compared with NA3 data~\cite{Badier:1982ae}.
Within the uncertainties, no significant evidence of nuclear effects is observed.

Extensive studies are performed to quantify the systematic uncertainty of the double $\Jpsi$ production cross section measurement. 
For the NH$_{3}$ case the main contributions come from the evaluation of single and double $\Jpsi$ acceptances; they are estimated to be 1.4 pb/nucleon and 2.5 pb/nucleon, respectively.
The uncertainty of double $\Jpsi$ acceptance also takes into account the uncertainty of relative contributions of $q\bar{q}$ and $gg$ used in Monte-Carlo simulations. Another significant contribution (1.4 pb/nucleon) is due to the uncertainty of single $\Jpsi$ production cross section measured by NA3~\cite{Badier:1983dg}. Several other sources of systematic uncertainties are studied and found to be negligible, e.g.: background estimation procedure, the uncertainty of the estimated number of single $\Jpsi$ events and the contribution from $\Jpsi$ particles produced in pileup events.
For the other two targets, the main contribution to the systematic uncertainties comes from the evaluation of the combinatorial background.

\begin{table}[htp]
	\caption{Number of single and double $\Jpsi$ (selected candidates, background and signal) events, acceptance values of single and double $\Jpsi$ and cross section of double $\Jpsi$ on the COMPASS targets.}
	\label{tab:Crosssections}
	\begin{center}
		\begin{tabular}{|c|c|c|c|}
			\hline
			& NH$_3$ & Al & W \\
			\hline
			$N_{\Jpsi}$/10$^6$, events & 6.23 & 0.46 & 2.51 \\
			$N_{2\Jpsi~candidates}$, events & 28 & 2 & 13 \\
			$N_{2\Jpsi~background}$, events & 2.9 $\pm$ 0.5 & 1.4 $\pm$ 0.4 & 8.5 $\pm$ 2.0 \\
			$N_{2\Jpsi}$, events & 25.1$\pm$0.5 & 0.6$\pm$0.4 & 4.5$\pm$2.0\\
			$A_{2\Jpsi}$ & 0.129 & 0.051 & 0.050 \\
			$A_{\Jpsi}$ & 0.194 & 0.074 & 0.066 \\ 
			$\sigma_{2\Jpsi}$, pb/nucleon & $10.7 \pm 2.3_{stat} \pm 3.2_{syst}$ & $3.6 \pm 8.2_{stat} \pm 1.4_{syst}$ & $3.3 \pm 3.0_{stat} \pm 1.8_{syst}$ \\
			\hline
		\end{tabular}
	\end{center}
\end{table}%

\begin{figure}
	\begin{center}
		\includegraphics[width=220px]{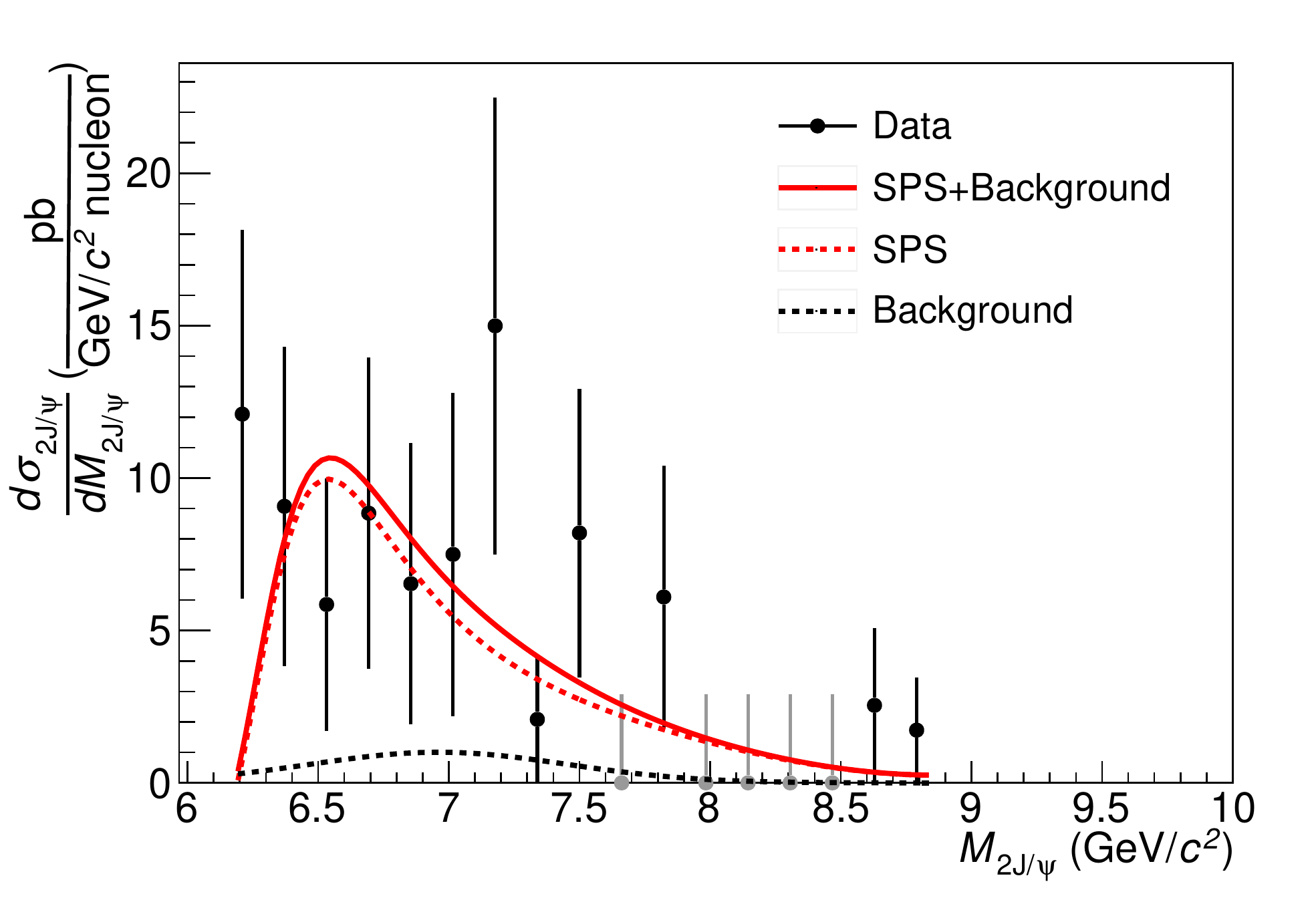}
		\includegraphics[width=220px]{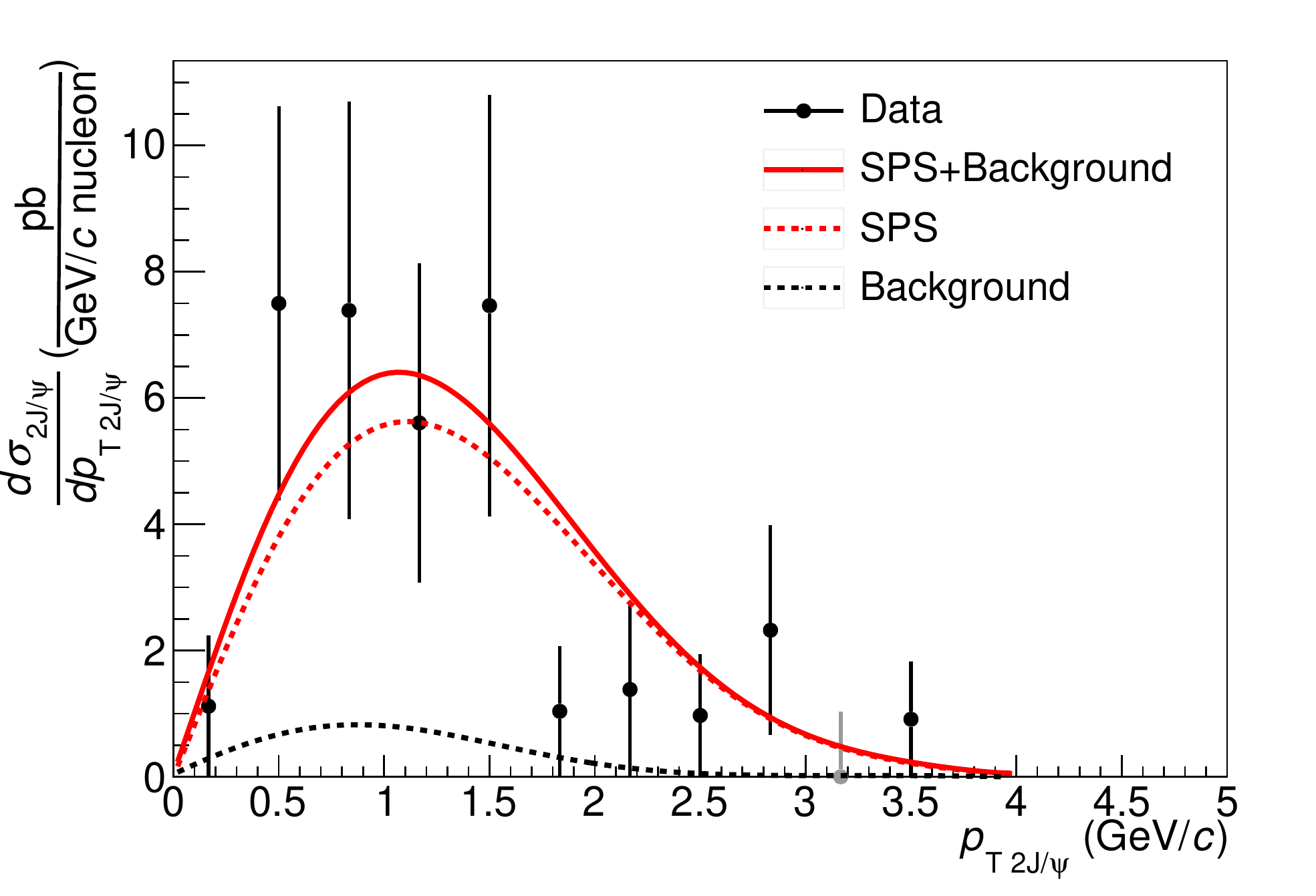}
		(a)\hspace{0.4\textwidth}(b)$\quad$ 
		
		\includegraphics[width=220px]{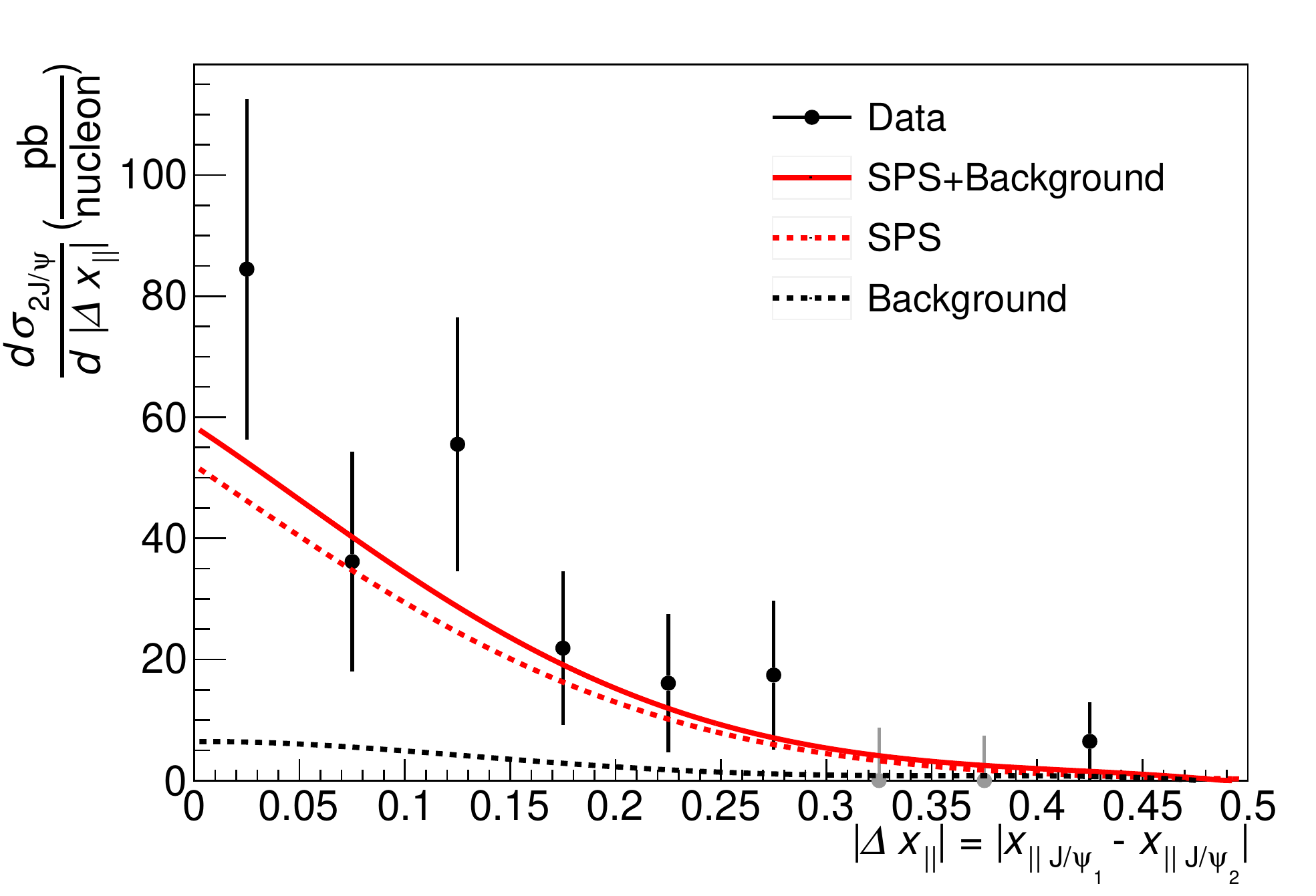}
		\includegraphics[width=220px]{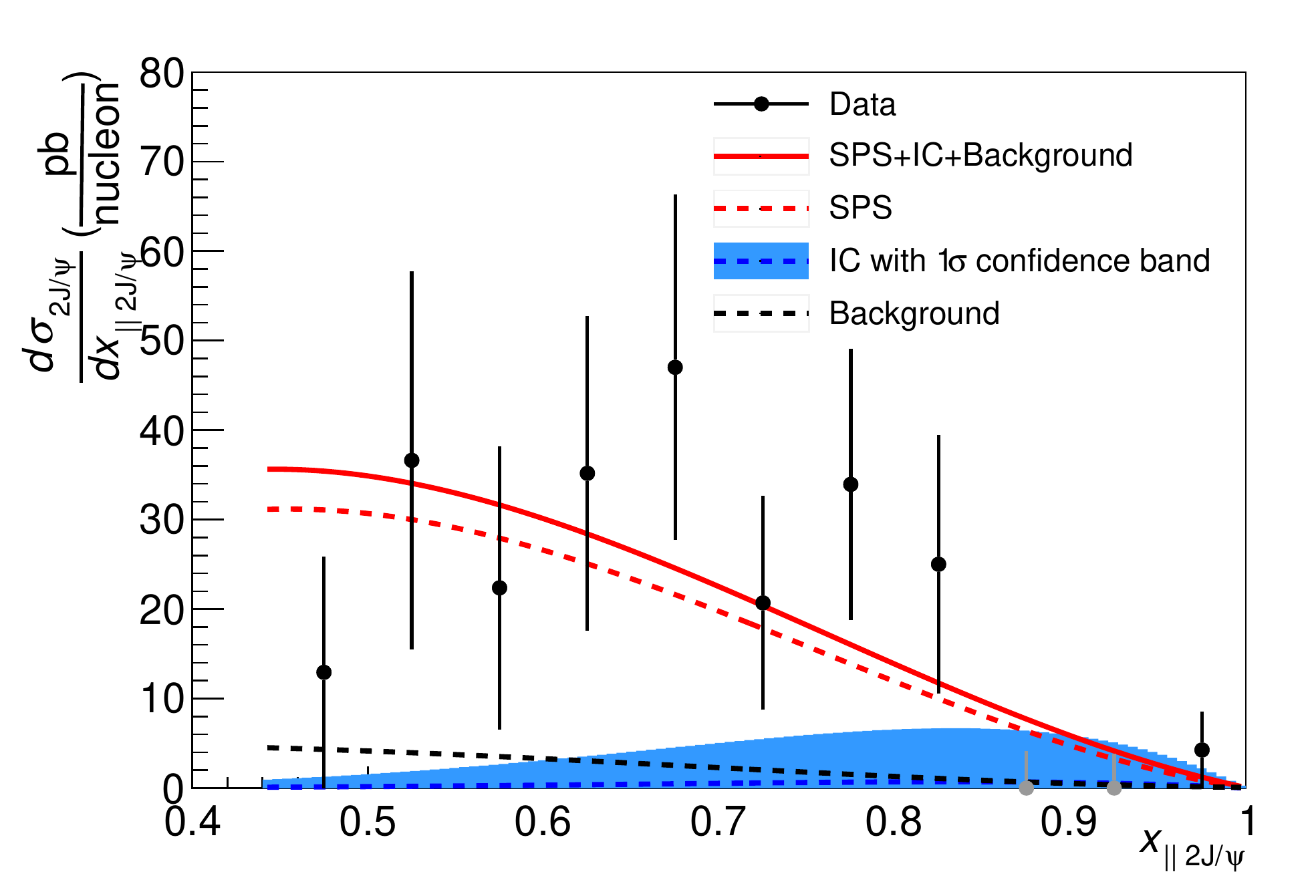} 
		(c)\hspace{0.4\textwidth}(d)$\quad$ 
	\end{center}
	\label{fig:fig4}
	\caption{
		(a) The double $\Jpsi$ production cross section per nucleon as a function of the invariant mass, $M_{2\Jpsi}$. The data points are compared to the sum of SPS and background contributions (solid red line). Individual SPS (dashed red line), and background (dashed black line) contributions are also shown.
		\newline
		(b) The double $\Jpsi$ production cross section per nucleon measured as a function of the transverse momentum, $\pt$. Curves are as described above for panel (a).
		\newline
		(c) The double $\Jpsi$ production cross section per nucleon as a function of the difference of the longitudinal momentum fractions of the two $\Jpsi$ mesons, $|\Delta x_{||}|$. Curves are as described above for panel (a).
		\newline
		(d) The cross section of double $\Jpsi$ production per nucleon as a function of the double $\Jpsi$ longitudinal momentum fraction, $\xpar$. The COMPASS data points are compared to the sum of SPS, IC and background contributions (solid red line). Individual SPS (dashed red line), background (dashed black line) and IC (blue dashed line) contributions are also shown. The IC contribution is shown with one standard deviation uncertainty band.}
\end{figure}

Due to low statistics and significant background contribution in the double $\Jpsi$ samples from Al and W targets, only the kinematic distributions from the ammonia target will be discussed in the following. The acceptance values evaluated for each event are used to calculate the average bin-by-bin acceptance for each kinematic distribution.
The acceptance-corrected distributions are normalised to the background-uncorrected integrated cross section.
The obtained differential cross sections are presented in Fig.~\ref{fig:fig4}.

The cross section as a function of the invariant mass of the two-$\Jpsi$ system is shown in Fig.~\ref{fig:fig4}(a). The red dashed curve illustrates the contribution of background events generated with Pythia 8~\cite{Sjostrand:2014zea} and is normalised to the integrated background estimated using values from Tab.~\ref{tab:Crosssections}.
The mass spectrum does not exhibit any statistically significant resonant structure.
The double $\Jpsi$ mass range corresponding to $\eta_b$ and $\chi_{b0,2}$ decays ($M_{\eta_b, \chi_{b0,2}} > 9$ GeV/$c^{2}$), which is referred to in the Introduction, is practically not accessible in the current measurement.

In Fig.~\ref{fig:fig4}(b) the differential cross section $d\sigma_{2\Jpsi}/d\pt$ is shown, where  $\pt$ is the transverse momentum of the double $\Jpsi$ system with respect to the beam track. 
The distribution extends up to $\pt\approx3.5$ GeV/c and the mean value is $\langle \pt \rangle$=1.3 $\GeV/c$.
The differential cross section as a function of $|\Delta x_{||}|=|x_{||~\Jpsi_1}-x_{||~\Jpsi_2}|$ is presented in Fig.~\ref{fig:fig4}(c), where $x_{||~\Jpsi}=p_{\mathrm{L}\Jpsi}/\pbeam$. Here $p_{\mathrm{L}\Jpsi}$ is the longitudinal momentum of $\Jpsi$ with respect to the pion beam direction in the target rest frame and $\pbeam$ is the pion momentum. 
The $\pt$ and $|\Delta x_{||}|$ distributions are in agreement with SPS model expectations, however within present statistics cannot be used to disentangle different production mechanisms~\cite{Gridin:2019nhc}.

The double $\Jpsi$ production cross section as a function of  $\xpar=\pz/\pbeam$ is presented in Fig.~\ref{fig:fig4}(d), where $\pz$ is the longitudinal momenta of the double $\Jpsi$ system defined along the pion direction in the target rest frame.
The function
\begin{equation}
	f(x_{||~2\Jpsi})=a\cdot \fsps + b\cdot \fic + \fbckg,
\end{equation}
where $a$ and $b$ are free parameters, is fitted to the experimental points in the kinematic range $\xpar > 0.44$. Here $\fsps$ represents the contribution of the SPS mechanism and $\fic$ corresponds to a possible contribution of the pion intrinsic charm. The SPS distribution is generated by the HELAC-Onia package and the IC parameterisation is taken from Ref.~\cite{Gridin:2019nhc}. The background contribution $\fbckg$ is generated using Pythia 8 and normalised  using the integrated values presented in Tab.~\ref{tab:Crosssections}.
The contribution from the DPS production mechanism is neglected, since it is estimated to be smaller than the background contribution and it is expected not to exceed 8\% of the SPS~\cite{Koshkarev:2019crs}. In contrast to IC, both DPS and SPS distributions are expected to peak at relatively small values of $\xpar$.
In Fig.~\ref{fig:fig4}d the fit result is represented by the solid red curve, while dashed black and red lines correspond to SPS and background contributions, respectively. The contribution of IC with one standard deviation uncertainty band is shown in blue. The experimental points are fully consistent with the SPS hypothesis, which appears to be sufficient to describe the data.
The upper limit on the possible contribution of the intrinsic charm mechanism to the integrated cross section is estimated to be
$\sigma_{IC}/\sigma_{2\Jpsi}<0.24~(CL=90\%)$.

\section{Discussion and summary}
\label{section:section5}
In the context of the COMPASS measurement, it is interesting to revise the interpretations~\cite{Vogt:1995tf,Koshkarev:2016rci,Brodsky:2017ntu,Koshkarev:2018xhd,Koshkarev:2017txl} of the NA3 result on the double $\Jpsi$ production~\cite{Badier:1982ae} and the SELEX results on the production of double charm baryons~\cite{SELEX:2002wqn,SELEX:2004lln}. These results were interpreted as an evidence for the intrinsic charm mechanism, however, neglecting other contributions (e.g. SPS). Although it is not possible to compare directly SELEX and NA3 results, it has been shown~\cite{Brodsky:2017ntu} that the ratios of integrated partonic production cross sections, $\sigma(c\bar cc\bar c)/\sigma(c\bar c)$, calculated for SELEX and NA3 are compatible within uncertainties. As it was shown in Section~\ref{section:section4}, the NA3 and COMPASS results are also compatible, whereas the latter appears to be mainly driven by the SPS mechanism.
Hence relying exclusively on the IC hypothesis to describe both NA3 and SELEX production rates is not justified and the SPS contribution can be the dominant one as it is the case for COMPASS.

In conclusion, the inclusive double $\Jpsi$ production is studied by the COMPASS experiment using a pion beam scattering off various nuclear targets. The differential cross section is measured as a function of $\mpair$, $\xpar$, $\pt$ and $\Delta x_{||}$.
No evidence of any resonant states decaying into two $\Jpsi$ is found within the limited statistics of this measurement.
To discriminate the leading production mechanism, the differential cross section $d\sigma_{2\Jpsi}/d\xpar$ is used since contributions from IC and SPS mechanisms are expected to peak in different $x_{||~2\Jpsi}$ regions.
Both SPS and IC hypotheses are used to fit $d\sigma_{2\Jpsi}/d\xpar$.
The upper limit on the production rate of double $\Jpsi$ from the intrinsic charm mechanism is estimated.
The obtained result for the differential cross section $d\sigma_{2\Jpsi}/d\xpar$ is fully consistent with the SPS hypothesis which appears to be sufficient to describe the data.
Within estimated uncertainties the contribution of intrinsic charm is found to be small and compatible with zero.

%%%%%%%%%%%%%%%%%%%%%%%%%%%

\section*{Acknowledgements}
We gratefully acknowledge the support of the CERN management and staff as well as the skills and efforts of the technicians of the collaborating institutions.
\newline
The work presented in this paper was done before February 2022.

%%%%%%%%%%%%

\end{document}